\documentclass[11pt]{article}

\begin{document}\thispagestyle{empty}
\begin{flushright}
\framebox{\small BRX-TH~542}\\
\end{flushright}

\vspace{.8cm} \setcounter{footnote}{0}

\begin{center}{\Large{Shortcuts to Spherically Symmetric Solutions: \\ A Cautionary Note}}\\[8mm]

S. Deser\\
Department of Physics, Brandeis University\\ Waltham, MA 02454,
USA \vspace{.4cm}

J. Franklin\\
Center for Space Research, MIT\\
Cambridge, MA 02139, USA \vspace{.4cm}

B. Tekin\\
Department of Physics, Middle East Technical University\\
06531, Ankara, Turkey \vspace{.4cm}

{\small(\today)}\\[1cm]
\end{center}

\begin{abstract}
\noindent Spherically symmetric solutions of generic gravitational models
are optimally, and legitimately, obtained by expressing the action in terms of the 
surviving metric components.  This shortcut is not to be overdone, however:  a one-function
ansatz invalidates it, as illustrated by the incorrect solutions of~\cite{Wohlfarth}. 
\end{abstract}

It is well-understood since the original work of Palais~\cite{Palais} that Êthe 
field equations describing highly symmetric geometries can be obtained legitimately
by using only candidate metrics endowed with those symmetries.  This simplifying
procedure does require some care, however, as illustrated in~\cite{DT}.  In particular, 
one must retain enough dependent variables in order to probe the 
required range
of variation of the action.  Specifically, spherically symmetric metrics have two independent
components ($g_{rr}$, $g_{00}$) in the natural Schwarzschild gauge.  While the 
field equations of, say Einstein gravity imply $g_{00} g_{rr} = Ê-1$, assuming this a priori in the
Lagrangian density turns it into a total divergence, with similar inconsistencies in other actions.  This
is just a special case 
of the generic~\footnote{The few cases that permit this process, such as removal of a 
constraint variable $F$,
show how restrictive it is: $F$ (but not $\phi$!) may be eliminated from 
$L=F^2 -2F X(\phi)$,
where $X$ is a function of $\phi$ and its derivatives, by completion of 
squares.}
invalidity of ``going on half-shell" in obtaining Euler-Lagrange 
equations.  Actually, there is a deep problem~\cite{DF} associated with even the Weyl $2$-function ansatz -- its inability to reproduce Birkhoff's theorem;  given the more immediate difficulties addressed below, we drop it here.

An example of this error is a recent work [1] in this Journal 
claiming 1-function spherical solutions to Born-Infeld gravity (BIG) models~\cite{DG}. 
It is not even necessary to delve into its numerical calculations; the 
assertion that its single field equation is solved exactly by any monomials $r^n$ ,
$n=0,1,2,3$, and that these solutions also solve the full field equations is already false: as is trivially checked, $r^3$ does {\it not} solve the original field equations. Neither $r^0$ nor $r^1$ are solutions to even the single field equation (14) because an arbitrarily 
discarded denominator vanishes there. Only $r^2$ , which represents flat space 
and does solve BIG, is a solution to both the 1-function and 2-function field equations (but only once the incorrect action (11) is properly rewritten~\footnote{The form of (11) does not even correspond to its 
antecedent (8)
because $\sqrt{-g}$ was improperly distributed. This leaves the equation 
(14) for $D(r)$ unchanged, but alters the ones for the original two components.}).

We need only conclude with the following immediate disproof:  BIG contains General Relativity as the first term in its expansion~\cite{DG}
\begin{equation}
I =\int d^4 x \left(\sqrt{-\vert g_{\mu \nu} + \lambda R_{\mu \nu} \vert} - \sqrt{-g} \right) \rightarrow \frac{\lambda}{2}\left( \int d^4 x \sqrt{-g} R\right)  + O(\lambda^2).
\end{equation}
But we have just seen that the ansatz of~\cite{Wohlfarth} misses this term altogether, since it becomes a total divergence (as demonstrated in~\cite{DT}).  
\section*{Acknowledgements}
This work was supported by NSF grants PHY99-73935 and PHY04-01667.

%\section*{References}

\bibliography{DFT.arxiv.ver2}

 \end{document}